\documentclass[prl,twocolumn]{revtex4} 
\usepackage{graphicx}
\usepackage{amsmath}
\usepackage{amsfonts}
\usepackage{mathrsfs}
\usepackage{color}
\usepackage{slashed}
\usepackage{comment}
\usepackage{epstopdf} %Aquest usepackage l'he de fer servir per compilar al meu portatil
\usepackage[normalem]{ulem}
\newcommand{\tn}{\textbf}

\def\<{\langle}
\def\>{\rangle}

\def\beq{\begin{equation}}
\def\eeq{\end{equation}}
\def\barray{\begin{eqnarray}}
\def\earray{\end{eqnarray}}

\newcommand{\be}{\begin{equation}}
\newcommand{\ee}{\end{equation}}

\font\numbers=cmss12
\font\upright=cmu10 scaled\magstep1
\def\stroke{\vrule height8pt width0.4pt depth-0.1pt}
\def\topfleck{\vrule height8pt width0.5pt depth-5.9pt}
\def\botfleck{\vrule height2pt width0.5pt depth0.1pt}
\def\Zmath{\vcenter{\hbox{\numbers\rlap{\rlap{Z}\kern
0.8pt\topfleck}\kern 2.2pt
                   \rlap Z\kern 6pt\botfleck\kern 1pt}}}
\def\Qmath{\vcenter{\hbox{\upright\rlap{\rlap{Q}\kern
                   3.8pt\stroke}\phantom{Q}}}}
\def\Nmath{\vcenter{\hbox{\upright\rlap{I}\kern 1.7pt N}}}
\def\Rmath{\vcenter{\hbox{\upright\rlap{I}\kern 1.5pt R}}}
\def\Pmath{\vcenter{\hbox{\upright\rlap{I}\kern 1.5pt P}}}

\begin{document}

%\date{\today}
\title{Tensor networks for frustrated systems: emergence of order from simplex entanglement}
\author{Daniel Alsina$^1$ and Jos\'e I. Latorre$^{1,2}$ \vspace{0.2cm} \\  
${}^1$ Departament d'Estructura i Constituents de la Mat\`eria, Universitat de Barcelona, 
%08028  
Barcelona, Spain  \\ 
 ${}^2$  
Center for Quantum Technologies, National University of Singapore, Singapore
}

\begin{abstract}
We consider a frustrated anti-ferromagnetic triangular lattice Hamiltonian and show that the properties of the manifold of its degenerated ground state are 
represented by a novel type of tensor networks. These tensor networks
are not based on ancillary maximally entangled pairs, but rather on triangular W-like simplices. 
Anti-ferromagnetic triangular frustration
is then  related to ancillary W-states in contrast to ferromagnetic order
which emerges from the contraction of GHZ-like triangular simplices.
We further discuss the outwards entangling power of various simplices. 
This analysis suggests the emergence
of distinct macroscopic types of order from the classification of entanglement residing on the simplices that define a tensor network. 
\end{abstract}

\maketitle

Tensor networks stand as a powerful variational approach to quantum mechanical systems that
escapes the sign problem that hampers Monte Carlo simulation \cite{sign}. Popular 
tensor networks are those that adapt their connectivity to the natural setting of a
particular system. Relevant instances of tensor networks include Matrix Product States (MPS) \cite{MPS} and Projected Entangled Pairs States 
(PEPS) \cite{PEPS} for translational invariant
systems in one  and two dimensions respectively, and MERA structures \cite{MERA} for 
scale invariant dynamics. All these approximation techniques are based on the use of ancillary
maximally entangled states that are projected in different manners along the network.
Yet, this strategy seems to capture the physics of frustration in a poor way. We here shall 
propose the construction of a novel type of tensor networks based on triangular
ancillary states which are related to W-type entanglement and, as a consequence,
are suited to describe geometric frustration.

\begin{figure}[h]
      \begin{center}
            \includegraphics[scale=0.27]{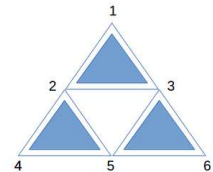}
            \caption{\label{fig:triangle}Example of triangular lattice geometry. Only up triangles have to be
considered so as not to double count each link.
             }
      \end{center}
      
\end{figure}

We first should note that the word frustration is used with different meanings in classical 
\cite{classicalfrustration} and
quantum physics \cite{quantumfrustration}.
For quantum systems, frustration often  describes the situation
where a Hamiltonian is made of terms that do not commute and, hence,
there is no global eigenstate that minimizes each of the terms in the Hamiltonian. 
The system then develops entanglement.
There is a different type of frustration which is found both in quantum and classical systems where the Hamiltonian is made out of commuting pieces, yet the state of minimum energy
does not minimize each term due to constraints emerging from geometry. 
The prototype example
of this kind of geometrical frustration is modeled by the anti-ferromagnetic 
triangular lattice model that we present in its quantum version
\begin{equation}
H= J \sum_{\{ i,j\} \in T} \sigma^z_i \sigma^z_j  + \lambda \sum_i \sigma^x_i ,
\label{eq:antifmham}
\end{equation}
where $\sigma^z$ and $\sigma^x$ are the Pauli matrices, $\{ i,j \} \in T$ represents nearest neighbor
interaction on the triangular geometry shown
in Fig. \ref{fig:triangle}, $J>0$ corresponds to the antiferromagnetic interaction and $\lambda$ is the external transverse field. It is easy to see that even in the case of $\lambda=0$ and $J>0$  
there is no arrangement of spins that minimize every term in the
Hamiltonian.

Let us recall that the classical counterpart of the anti-ferromagnetic model displays a large degeneracy of ground states. 
In a triangular lattice made of $n$ spins and $\lambda=0$, Wannier found that the number of states $M$ with the same minimum energy $E_0=-n/3$ 
 scales as the volume
of the system, that is $M\sim 2^{\alpha n}$, where $\alpha \sim .488$ \cite{Wannier}. 
This large degeneracy translates to the
quantum case in the form of a large subspace of possible ground states $| \psi_i\rangle$,
$i=1,\ldots,M$, 
where each $|\psi_i\rangle$ corresponds to the quantum transcription of a classical solution
into the computational basis. We may then
consider the equal superpositions of all valid states on the computational
basis that carry the same minimum energy $E_0=-n/3$,
\begin{equation}
  \vert \psi_0\rangle =\frac{1}{\sqrt M}\sum_i \vert \psi_i\rangle \qquad H |\psi_i\rangle=E_0 
|\psi_i\rangle , \quad i=1\ldots,M
\end{equation}
where $M$ corresponds to the degeneracy of the ground state manifold.
The degeneracy is lifted
when an external transverse field is applied. 
The special
case where ${\lambda\to 0}$ produces a particular combination of all $|\psi_i\rangle$.
In both cases, namely in the equal superposition or in the limit to zero of the 
external field, the resulting states carry a large entropy, as we shall discuss later.

The question we shall address here is whether a frustrated state such as $|\psi_0\rangle$ accepts a natural representation in terms of tensor networks.
Let us start by observing than in the case of 3 spins forming a triangle, the
equal superposition of valid ground states is
\begin{eqnarray}
\nonumber
  | \psi_0 \rangle &&= \frac{1}{\sqrt 6} \left(|001\rangle+|010\rangle+|100\rangle \right. \\
  && \left. +|011\rangle+|101\rangle+|110\rangle \right) \nonumber \\  && = \frac{1}{\sqrt 2}
 \left( | W \rangle + | \bar W \rangle \right) ,
\end{eqnarray}
where $| W\rangle = \frac{1}{\sqrt 3}\left(|001\rangle+|010\rangle+|100\rangle\right)$ and $| \bar W\rangle = \frac{1}{\sqrt 3}\left(|011\rangle+|101\rangle+|110\rangle\right)$. This introduces in a natural way  $W$-states which are attached to
a class of tripartite entanglement in quantum information theory \cite{W}. Frustration in this state amounts
to equally superpose all the possibilities of assigning frustrated triangles, 
hence the emergence of $W$-type entanglement. Note that if we relax
the requirement of equal superposition of possible ground states but retain
isotropy, a freedom on the relative weight of $| W\rangle$ {\sl vs.} $| \bar W \rangle$
states appears. 

We next consider a larger structure, as the one shown in Fig. \ref{fig:triangle}, and 
compute the ground state in the limit of zero external transverse field.
The result reads
\begin{eqnarray}
\nonumber
 | \psi_0 \rangle &&= \alpha \left(|001100\rangle+|010001\rangle+|011101\rangle\right)
\\ &&+ \beta \left(|001101\rangle+|010011\rangle+|001110\rangle \right.+
\nonumber
\\ &&\left.|010101\rangle+|011001\rangle+|011100\rangle\right)
\nonumber
\\ &&+ \gamma \left(|001010\rangle+|010010\rangle+|011000\rangle\right) \nonumber
\\ &&+ \delta |011010\rangle \ .
\label{ground6}
\end{eqnarray}
with $\alpha\sim -.24$, $\beta\sim .19$, $\gamma\sim -.16$ and $\delta\sim .15$.
It is convenient to analyze this state by looking at the distribution of spins
on all triangles pointing up, since the triangles pointing down only provide
a redundant description of the system. There are three up-triangles to be considered,
respectively formed by the qubits \{1,2,3\}, \{3,4,5\} and \{3,5,6\}.
The relevant observation is that each state forming the superposition of
the ground state in Eq. \ref{ground6} is formed by a member of either a $W$ state or a $\bar W$ state.
It is furthermore possible to verify that all the
states in the equal superposition of valid ground states are made of elements of $W$- and
$\bar W$-like states in the up-triangles.
Though we may find a down-triangle with the configuration
111, this does not invalidate the fact that all up-triangles remain a member of genuine $W$ tripartite entanglement. 

Let us now prove that this is the general case, namely that
the equal superposition of valid ground states $| \psi_0\rangle$
is made of all possible combinations of $W$ and $\bar W$ configurations
on all ancillary simplices. To prove this result we first 
consider the Hamiltonian
\begin{equation}
  H_W=\sum_{i,j,k\in T^*} (z_i+z_j+z_k-1)^2 ,
\end{equation}
where $z_i=\frac{1+\sigma^z_i}{2}$  and $T^*$ spans the set
of up-triangles. Expanding this Hamiltonian we find
\begin{equation}
  H_W=\frac{1}{2}\sum_{i,j\in T} \sigma^z_i \sigma^z_j + \sum_i \sigma^z_i + 1 ,
  \label{eq:wham}
\end{equation}
This construction shows that the ground state of the Hamiltonian $H_W$ belongs to
the manifold spanned by $W$ states and that it carries $E_0=0$ energy.

This intuitive technique to construct frustrated Hamiltonians indicates
that the way to represent the frustrated triangular dynamics 
needs to cancel the linear terms in $\sigma^z$.
This can be done as follows
\begin{eqnarray}
  \nonumber
  H&&=\sum_{i,j\in T} \sigma^z_i \sigma^z_j \\
  \nonumber
&&=
  \sum_{i,j,k\in T^*} \left( (z_i+z_j+z_k-1)^2 \right.\\
&&\left. + (z_i+z_j+z_k-2)^2 -2 \right) \ ,
\end{eqnarray}
where now ${i,j,k}$ are indexes of the sites that form up-triangles $T^ *$.
The original anti-ferromagnetic triangular Hamiltonian is recovered as a
sum of two conditions, one trying to produce a superposition of $W$ states and another
of $\bar W$ states. Both conditions can not be fulfilled simultaneously,
so that the energy picks a penalization of one unit for each up-triangle coming
from one of the two terms, 
hence $E_0=-n/3$, which is the number of up-triangles.

\begin{figure}[h]
      \begin{center}
            \includegraphics[scale=0.22]{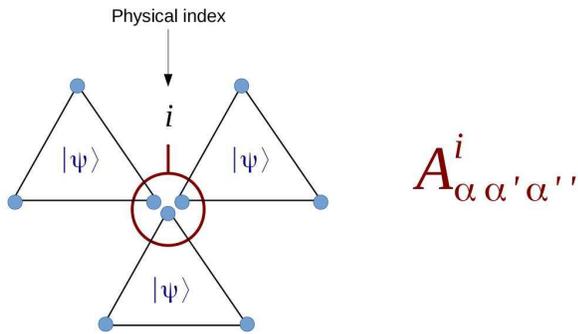}
            \caption{\label{fig:tensor}Detail of the way three concurrent qubits with
            indexes $\alpha$, $\beta$ and $\gamma$, coming from
 entangled triangles are projected into a physical index $i$ defining the
tensor $A^i_{\alpha\beta\gamma}$. The triangular anti-ferromagnetic 
equal superposition of ground states is obtained if $|\psi\rangle=
\frac{1}{\sqrt 2}\left( | W \rangle + | \bar W \rangle \right)$ and $A^i_{\alpha\beta\gamma}=\delta^i_\alpha \delta^i_\beta \delta^i_\gamma$.
             }
      \end{center}
      
\end{figure}

The above arguments allow for the construction of a tensor network that
represents the state $| \psi_0\rangle$ in an exact way. We first consider
the filing of up-triangles all across the triangular lattice with $W$ and
$\bar W$ ancillary states, that is ancillary quantum degrees of freedom
of dimension $\chi=2$,
 as shown in Fig. \ref{fig:tensor}.

\begin{figure}[h]
      \begin{center}
            \includegraphics[scale=0.27]{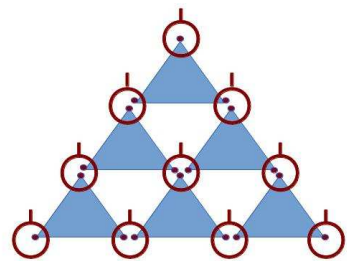}
            \caption{\label{fig:tensornetwork}Tensor network based on a triangular simplex described 
by the tensor $A^ i_{\alpha\beta\gamma}$,  which projects the underlying ancillary indexes onto
a physical one.
             }
      \end{center}
      
\end{figure}

%\begin{comment}
%\end{comment}

 We then consider the projection
\begin{equation}
A^i_{\alpha\beta\gamma}=\delta^i_\alpha \delta^i_\beta \delta^i_\gamma .
\end{equation}
That is, the tensor only takes non-vanishing +1 value if all ancillary indexes 
$\alpha, \beta,\gamma$ agree at
that point, and pass their value to the physical index $i$.
It is easy to check that the contraction of ancillary indexes for this tensor network 
reproduces the state $| \psi_0\rangle$. Notably, a similar construction
based on the use of $GHZ$ states at each simplex does describe global ferromagnetic
order. The interplay between $GHZ$ and $W$ entanglement at the level of 
ancillary simplices is the key to distinct types of order at large scales.

A first consequence from the above construction is to observe that
the entanglement of the state $|\psi_0\rangle$ is bounded to obey
area law scaling at most. 
As mentioned previously,  $|\psi_0\rangle$
is made of an exponential superposition of states. Thus, in principle, this
state could display a volume law scaling of entanglement.
Yet, $|\psi_0\rangle$ is
described by a local tensor network that only links each spin to its nearest neighbor.
That immediately sets a bound on the entropy of the state. Indeed, the entanglement
entropy of the state $| \psi_0\rangle$ will scale as the area law at most. 
Moreover the ancillary states carry dimension $\chi=2$, so the bound for
the entanglement entropy of a region A with boundary $\partial A$ is just $S(A)\le
\log 3 \ \partial A$, being 3 the options that emerge outwards from each qubit. 
This is fully consistent with the idea that local interactions of translational
invariant systems produce ground states that obey area law scaling for the entanglement
entropy \cite{arealaw}.

Let us now show that the triangular frustrated system is a particular case of 
the Exact Cover NP-complete problem, closely related to 3-SAT problem. Exact Cover
is a decision problem based on the fullfilment of 3-bit clauses.
To be precise, we are asked  to decide whether a set of $n$ bits
accept an assignment such that a set of clauses involving three bits are all satisfied.
Each clause is obeyed if the three bits involved in it take values 001, 010 or 100.
It was proven in Ref. \cite{exactcover} that there is an exact tensor network that describes the 
possible solutions of this problem and that the hard part
of deciding the instance is found in the contraction of the tensor network. 

In our case, the construction we have
proposed previously can be seen as the particularization of the 
Exact Cover to the problem of a regular triangular lattice.
This implies that triangular frustration is a sort of simple and regular version of Exact Cover.
Indeed, Exact Cover clauses involve bits that have no geometrical proximity relation. 

This observation can be translated to a statement about frustration cycles.
The triangular model produces frustration at the level of single triangles.
Instead,  Exact Cover produces frustration over a  non-local 
and non-homogeneous triangular lattice. Typical cycles of frustration in
Exact Cover are of $\log n$ size: therefore,  the NP  Exact Cover 
problem is much harder than regular triangular
lattices models because of the long scale cycles for frustration.

\begin{figure}[h]
      \begin{center}
            \includegraphics[scale=0.27]{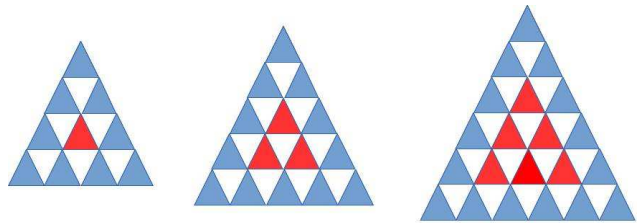}
            \caption{\label{fig:outwards}Outwards entangling power of simplices can be
assessed by computing the entropy of the reduced density matrix corresponding to the
red (or dark) area made of $n_A$ spins, with $n_A = 3, 6, 10$ in this figure.
             }
      \end{center}
     
\end{figure}

We now turn to the issue of how ancillary entanglement  develops into large scale correlations.
We shall refer to this property as {\sl outwards entangling power} of an ancillary simplex.
We may visualize this process by first focusing on a single ancillary triangle.
The different superpositions which are accepted on this ancillary state 
propagate outwards distinct configurations. For instance, the trivial case where
the internal state corresponds to a 000 configuration can only propagate a global
product state made of zeros. In the case where a $W$ ancillary state is used,
the global state retains only a small amount of entanglement due to the dominant
role of 0 versus 1 ancillary states. A remarkable 
jump in entanglement entropy is obtained when $W$ and $\bar W$ are accepted at
the ancillary level. Then the global state gains a complex structure, and 
entanglement appears to scale. The outwards growth of entanglement from
an ancillary triangle can be systematically analyzed. In Fig. \ref{fig:outwards} and Table \ref{table:outwards} we 
present how much entanglement is observed as the size of the system increases.

%\begin{comment}
%\end{comment}

\begin{table}
  \begin{tabular}{ | c | c  c  c  |}
    \hline
      Simplex & $n_A=3$ & $n_A=6$ & $n_A=10$ \\ \hline
    $|GHZ\rangle$& 1 & 1 & 1\\
    $|W\rangle$ & $\log_2 3$ & $\log_2 3$ & $\log_2 3$ \\
    $|W\rangle$,$|111\rangle$ & 2 & 3&4 \\
    $|W\rangle$,$|\bar W\rangle$ & 2.183 & 3.126 & 5.053 \\
    	$|W\rangle$,$|\bar W\rangle$, $|111\rangle$ & 1.815 & 2.756 & 4.314 \\
   \hline 
  \end{tabular}
  \caption{Outward entanglement power of different triangular simplices. At each simplex,
an equal superposition of its allowed states (on the left) propagates the entanglement entropy (on the right, in ebits)
to the rest of the system.}
  \label{table:outwards}
\end{table}

The relation between $GHZ$ and $W$ entanglement on triangular ancillary states and
the emergence of distinct types of long distance order suggest interesting
generalizations for larger simplices.

\begin{figure}[h]
      \begin{center}
            \includegraphics[scale=0.18]{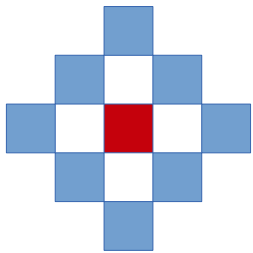}
            \caption{\label{fig:squares}Square network of 24 spins, with the red (or dark) area showing the 4 spins from which we are computing the entanglement entropy.
             }
      \end{center}
     
\end{figure}

Let us here analyze the case of a tensor network created from a four-qubit simplex.
We shall constrain our analysis to simplices that are symmetric under the exchange of
ancillary particles, namely
\begin{eqnarray}
  |\psi\rangle &&= \alpha_{0} |0000\rangle
  \nonumber
\\
&&
+ \alpha_1 \left(|0001\rangle +|0010\rangle +|0100\rangle +|1000\rangle \right) 
  \nonumber
\\
&&
+ \alpha_2 \left(|0011\rangle +|0101\rangle +|0110\rangle + \right. \nonumber 
\\
&& \left.|1001\rangle +|1010\rangle +|1100\rangle \right)
  \nonumber
\\
&&
+\alpha_3 \left(|0111\rangle +|1011\rangle +|1101\rangle +|1110\rangle  \right) \nonumber
\\
&& + \alpha_4 |1111\rangle \ .
\end{eqnarray}
We then consider a tensor network as shown in Fig. \ref{fig:squares} , where every checked square
contains an ancillary state and all links are counted just once. The tensor defined
at every site is the product of delta functions of the physical index with each of the
two ancillary qubits meeting there. 
For the case of 24 qubits shown in Fig. \ref{fig:squares} , we have scanned the entropy of the inner square as a function of the coefficients of this ancillary state.  Maximum
entanglement is obtained for the state corresponding to $\alpha_0=\alpha_2=\alpha_3=\alpha_4= - \alpha_1$.

%\begin{comment}
%\end{comment} 

The above results suggest a connection between microscopic entanglement at
the level of a simplex and the emerging of long-range correlations in the system.
It is tempting  to argue that distinct classes of entanglement might
be responsible for different types of long-range order, as we found in the case of the
triangular lattice.
For four qubits, it is known that there are nine classes
of 4-qubit states under the action of LOCC operations \cite{9classes}. There are 
the corresponding analogues
of $GHZ$ and $W$ states with four qubits, but there are further novel ways to entangle them.
States with maximal hyperdeterminant may play a special role in 3D networks based
on tetrahedrons.

Similarly, the spatial symmetries which are found on a simplex are related to symmetries 
at large scales. It is easy to see that if the triangular couplings are chosen as
positive in the diagonal directions and negative in the horizontal direction, then frustration
disappears and the exact simplex describing the model is a superposition of 
$|001\rangle$ and $|110\rangle$ states.

Therefore, we conclude that tensor networks constructed from non-trivial simplices are an interesting tool and could be the natural way to encode 
different levels of entanglement and symmetries at large order.

\indent \tn{Acknowledgements.} 
The authors are grateful to S. Iblisdir and A. Garc\'ia-S\'aez for their many very useful comments.
JIL acknowledges  financial support  from FIS2011-16185, Grup de Recerca Consolidat 
ICREA-ACAD\`EMIA, and National Research Foundation \& Ministry of Education, Singapore.

\end{document}